\documentclass[floatfix,aps,twocolumn,superscriptaddress,pra,10pt]{revtex4-1}

\usepackage{amssymb}
\usepackage{amsmath}
\usepackage[dvips]{graphicx}
\usepackage{bm}
\usepackage{float}
\usepackage{color}
	\usepackage{tabularx}

\graphicspath{{./figs/}}

\newcommand{\beq}{\begin{equation}}
\newcommand{\eeq}{\end{equation}}
\newcommand{\bea}{\begin{eqnarray}}
\newcommand{\eea}{\end{eqnarray}}

\newcommand{\up}{\uparrow}
\newcommand{\dn}{\downarrow}

\newcommand{\al}{\alpha}
\newcommand{\be}{\beta}

\begin{document}

\title{On the nature of the Mott transition in multiorbital systems}
\author{Jorge I. Facio}
\affiliation{Centro At{\'o}mico Bariloche and Instituto Balseiro, CNEA, CONICET, (8400) Bariloche, Argentina}
\author{V. Vildosola}
\affiliation{Departamento de Materia Condensada, GIyA, CNEA, CONICET, (1650) San Mart\'{\i}n, Provincia de Buenos Aires, Argentina}
\author{D. J. Garc\'ia}
\affiliation{Centro At{\'o}mico Bariloche and Instituto Balseiro, CNEA, CONICET, (8400) Bariloche, Argentina}
\author{Pablo S. Cornaglia}
\affiliation{Centro At{\'o}mico Bariloche and Instituto Balseiro, CNEA, CONICET, (8400) Bariloche, Argentina}

\begin{abstract}
We analyze the nature of Mott metal-insulator transition in multiorbital systems using dynamical mean-field theory (DMFT). 
The auxiliary multiorbital quantum impurity problem is solved using continuous time quantum Monte Carlo (CTQMC) and the rotationally invariant slave-boson (RISB) mean field approximation. 
We focus our analysis on the Kanamori Hamiltonian and find that there are two markedly different regimes determined by the nature of the lowest energy excitations of the atomic Hamiltonian. 
The RISB results at $T\to0$ suggest the following rule of thumb for the order of the transition at zero temperature: a second order transition is to be expected if the lowest lying excitations of the atomic Hamiltonian are charge excitations, while the transition tends to be first order if the lowest lying excitations are in the same charge sector as the atomic ground state. 
At finite temperatures the transition is first order and its strength, as measured e.g. by the jump in the quasiparticle weight at the transition, is stronger in the parameter regime where the RISB method predicts a first order transition at zero temperature. Interestingly, these results seem to apply to a wide variety of models and parameter regimes. 
\end{abstract}

\maketitle
\section{Introduction}

The transition between a Fermi liquid and a paramagnetic Mott insulator remains one of the most interesting phenomena driven by electronic correlations. When the interactions among electrons prevail over their itineracy there is an increase of the electronic effective masses and of the magnetic correlations. These effects give rise to fascinating properties, for instance, they are thought to be deeply related to the mechanism behind unconventional superconductivity like the one in Copper and Iron based materials \cite{Capone2364,PhysRevLett.93.047001,PhysRevB.91.085108}.

In his original analysis of the insulating character of nickel oxide, Mott pointed to the role of Coulomb interactions and argued, starting from the insulating phase, that there should be a sharp insulator to metal transition as the lattice spacing is reduced \cite{mott1949}. Hubbard introduced a lattice model with a single level per atom, a local repulsion $U$ and a hopping integral $t$ between nearest neighbor sites \cite{Hubbard238,hubbard1964electron}. At half-filling and low enough $t$ the system is in an insulating phase with upper and lower Hubbard bands, separated by a gap, which are associated to the dispersion of an extra electron or hole in the system, respectively. In the Hubbard picture the transition to a metal, as $t$ is increased, is expected to occur when the gap for charge excitations vanishes, i.e., when the average bandwidth of the upper and lower Hubbard bands is of the order of $U$.
A complementary analysis starting from the Fermi gas, was provided by variational methods like the Gutzwiller approximation \cite{PhysRevLett.10.159,PhysRev.137.A1726}.
These in turn gave place to the so-called Brinkman and Rice scenario of the metal-insulator transition (MIT) \cite{PhysRevB.2.4302,kotliar1984new},
in which as the Coulomb repulsion is increased, the effective mass of the low energy quasiparticles increases and diverges at the transition.

A bridge between these two limits was provided by the Dynamical Mean Field Theory (DMFT) 
\cite{georges1996dynamical} which was first used to analyze the transition between a paramagnetic metal and a paramagnetic insulator within the Hubbard model \cite{PhysRevLett.69.1236,PhysRevB.48.7167,PhysRevB.49.10181}.
The DMFT approximation made possible to treat on equal footing high energy features, as the Hubbard bands, 
and the low energy quasiparticle physics across the transition. 

A decade of studies \cite{PhysRevLett.69.1236,PhysRevLett.69.1240,PhysRevB.48.7167,PhysRevB.49.10181,PhysRevLett.74.2082,PhysRevLett.82.1915,PhysRevLett.82.4890,bulla1999zero,PhysRevLett.83.3498} of the phase diagram for the one band Hubbard model concluded that the transition 
is first order at finite temperatures with a critical second order end point. At zero temperature, 
the transition from the metallic side to the insulator occurs by a 
continuos reduction of the width of a quasiparticle band which, close to the MIT, is located between well separated Hubbard bands.

The experimental evidence of the Mott transition predicted originally for the
transition metal monoxides (TMMO) as NiO, MnO or FeO under pressure was evasive during more than five decades due to the high critical pressures required. 
Once confirmed \cite{PhysRevB.69.220101}, it proved to be a much richer phenomenon than expected due to the multiorbital physics of the 3d  band in the TMMOs \cite{PhysRevLett.94.115502,kunevs2008collapse}. 
The MIT in 4d materials, for which a multiorbital description becomes essential, has also been experimentally detected, such as in Ca2RuO4 driven by temperature \cite{PhysRevB.60.R8422}, pressure \cite{PhysRevB.65.220402} and doping \cite{PhysRevLett.84.2666}.
This is just to mention a few cases among the large variety of experiments that have exhibited signatures of a MIT in multiorbital systems \cite{Qazilbash1750,fujimori2012spectroscopy, PhysRevB.87.245109,PhysRevB.92.115142}. 

Recent experiments by Camjayi {\it et al.} \cite{camjayi2014} show clear indications of a first order phase transition in the GaTa$_4$Se$_8$ compound which can be modeled with a three orbital Hamiltonian. In the coexistence regime an external current can take the system from the insulating to the metallic phase and vice-versa, giving the possibility to use this compound as a resistive memory \cite{cario2010electric,stoliar2013universal}.

In the last few years, much progress has been achieved in understanding the role of interorbital interactions
in the electronic effective mass in multiorbital models and materials \cite{georges2013}. 
In general, the interorbital repulsion $U^\prime$ and the Hund's coupling $J$ modify the local multiplet structure in two ways:
by changing the energy gaps between multiplets in different charge sectors,
and by splitting multiplets within each charge sector, which changes the degeneracy of the atomic levels.
The level degeneracy is naturally of great importance within DMFT because the Kondo scale in the associated quantum impurity problem depends exponentially on it~\cite{schrieffer1967}.
The gap for charge excitations in the atomic limit is determined by the multiplet structure and is of crucial importance to set the critical value of the interactions that induce the MIT in the Hubbard picture.
In Refs. \cite{PhysRevLett.107.256401} and \cite{PhysRevB.83.205112} it was shown that these effects help to understand the 
antagonistic consequences of the Hund's coupling $J$ which, for some electronic fillings, increases both the effective mass of the quasiparticles and the critical interaction.

Interorbital couplings have also been reported to affect the way in which the quasiparticle band vanishes at the MIT.
B\"unemann $et$ $al.$ \cite{PhysRevB.55.4011,PhysRevB.57.6896} in a 
two-orbital implementation of the Gutzwiller approximation showed that the inclusion of $J$ modifies the Brinkman-Rice scenario
and that at half-filling the transition is first order at zero temperature, while it remains continuous for an average occupation of a single electron per site.
A similar effect of $J$ at half-filling was reported for two orbital models using different approximations to solve the DMFT equations \cite{PhysRevB.67.035119,hallberg2015sota,Prushke2005}.

Despite these important advances in the understanding of the multiorbital physics, a detailed study of the role of interorbital interactions in the nature of the MIT is still lacking. Precisely, how they affect the order of the transition, the quasiparticle weight and the way it vanishes at the MIT.

Here we report DMFT results for two and three orbital models at different electronic fillings. In order to disentangle the role of the interorbital interactions on the Mott transition we consider no crystal field splitting terms nor any asymmetry in the width or shape of the bands. We solve the DMFT equations using different quantum impurity solvers. The main results are obtained using the Rotationally Invariant Slave Bosons technique (RISB) in the mean field approximation \cite{Lechermann2007}. We also use the numerically exact Continuous Time Quantum Monte Carlo (CTQMC) at finite temperatures \cite{RevModPhys.83.349,triqs_ctqmc_solver_werner1,triqs_ctqmc_solver_werner2,triqs_ctqmc_gull}. 
Our RISB results in the $T\to0$ limit, based on calculations of the quasiparticle weight and of the lattice free energy, show that in multiorbital models the order of the transition at zero temperature in general depends on the electronic filling and on the values of interorbital interactions. We argue that this behavior can be understood in terms of the effects of interorbital interactions on the degeneracy of local multiplets and on the gap for charge excitations.

The rest of this paper is organized as follows. Section \ref{sec:methods} describes the model and the methods. In Sec. \ref{sec:model_u_up} the Mott transition is analyzed in the limit of vanishing Hund's rule coupling $J=0$. This simplified case with intraorbital ($U$) and interorbital ($U^\prime$) interactions contains the main ingredients needed to understand the physics of the more physically relevant case, with $J\neq 0$ and $U^\prime=U-2J$, which is treated in Sec. \ref{sec:model_K}. Finally, the main results concerning the nature of the MIT and the role of interorbital interactions determining it are summarized in Sec. \ref{sec:conclusions}.

\section{Models and Methods}\label{sec:methods}
We consider the Kanamori Hamiltonian to describe the local interactions in multiorbital systems:
\begin{equation}
	H=\sum_{ijmm^\prime\sigma}t_{ij}^{mm^\prime}d^\dagger_{im\sigma}d^{}_{jm^\prime\sigma}+\sum_i H_{i}^{\text{at}}
	\label{eq:hamtot}
\end{equation}
where $t_{ij}^{mm^\prime}$ is a hopping term between orbital $m$ on site $i$ and orbital $m^\prime$ on site $j$, and the local Hamiltonian $H_{i}^{\text{at}}$ is given for each site $i$ of the lattice by
\begin{eqnarray} \label{eq:kanamori}
	H^{\text{at}} &=& U \sum_{\al} n_{\al\up} n_{\al\dn} + U^\prime \sum_{\al\neq\be} n_{\al\up} n_{\be\dn} + \nonumber\\
&+& (U^\prime-J) \sum_{\al>\be,\sigma}n_{\al\sigma} n_{\be\sigma} - J \sum_{\al\neq\be} d^\dag_{\al\up} d_{\al\dn} d^\dag_{\be\dn} d_{\be\up} \\
&+& J \sum_{\al\neq\be} d^\dag_{\al\up} d^\dag_{\al\dn} d_{\be\dn} d_{\be\up}-\mu \hat{N}.\nonumber
\end{eqnarray}
Here $\hat{N} = \sum_{\al}\sum_{\sigma=\uparrow,\downarrow} n_{\al\sigma}$, $\mu$ is the chemical potential, $U$ and $U^\prime$ are the intraorbital and interorbital interactions, respectively, and $J$ is the Hund's rule coupling. 
As mentioned before, we focus our analysis on the role of the multiorbital interactions on the Mott transition. To that aim, we consider no crystal field splitting terms nor asymmetries in the width or shape of the bands, and set the interorbital hybridizations to zero. For simplicity we consider a semicircular density of states for each orbital:
\begin{equation}
	D(\varepsilon)=\frac{2}{\pi D}\sqrt{1-(\varepsilon/D)^2},
	\label{eq:ldos}
\end{equation}
where $D$ is the half-bandwidth of the conduction electron band in the absence of interactions, but our main conclusions do not depend on this choice.

We solved this model using DMFT \cite{georges1996dynamical}. This theory is based on the assumption of a local self-energy and maps the lattice interacting problem
onto a multiorbital quantum impurity problem where the impurity is described by $H^{\text{at}}$ and the electronic bath is subject to a self-consistency condition.

We implemented the Rotationally Invariant Slave Boson \cite{Lechermann2007} (see Ref. \cite{Bunemann1998} for a related approximation) in a quantum impurity formulation \cite{Ferrero2009,*Ferrero2009a}. 
The RISB formalism is a multiorbital generalization of Kotliar-Ruckenstein \cite{kotliar1984new} approach that preserves the rotational invariance at the mean field level \cite{li1989spin}. 
In this approximation the local electron operators $d_\al$, where $\alpha$ is an orbital index, are represented as a linear form in introduced quasiparticle operators $f_\beta$:
\begin{equation}
d_\al = R_{\al\be} f_\be.
\end{equation}
Here $R_{\al\be}$ depends on a set of parameters that need to be calculated minimizing a free energy.
The resulting self energy has a simple linear form, which in matrix notation reads:
\begin{equation}
	\mathbf{\Sigma}(i\omega_n)=i\omega_n \left( {\mathbf{1}-\mathbf{Z}^{-1}}  \right)+ \mathbf{R}^{\dag-1}\mathbf{\Lambda}\mathbf{R}^{-1} - \mathbf{\bm\epsilon_0},
	\label{eq:sigmaRISB}
\end{equation}
where $\mathbf{Z}$ is the quasiparticle weight which can be calculated as $\mathbf{Z} = \mathbf{R}\mathbf{R}^\dag$, 
$\mathbf{\bm\epsilon_0}$ contains the quadratic part of the atomic Hamiltonian and
$\mathbf{\Lambda}$ is formed by Lagrange multipliers introduced in order to enforce a proper mapping between
the original Hilbert space and its new representation.

We also solved the DMFT equations at finite temperatures using the numerically exact CTQMC impurity solver, for which
we use the TRIQS code \cite{Seth2015,Parcollet2015398}. For each impurity problem we performed typically 
$2\times 10^7$ measurements separated by 200 moves. We estimate the quasiparticle weight at finite temperatures as $Z_{\al\sigma} = [1-\text{Im}\Sigma_{\al\sigma}(i\omega_0) / \omega_0]^{-1}$.

\section{$J=0$ limit}\label{sec:model_u_up}

In this section we focus our analysis on the $J=0$ limit of the atomic Hamiltonian of Eq.(\ref{eq:kanamori}), which for $M$ orbitals, reads:
\begin{equation}\label{eq:h_up}
H^{\text{at}}_{J=0} = U \sum_{\al=1}^M n_{\al\up} n_{\al\dn} + U' \sum_{\al>\be,\sigma \sigma'}n_{\al\sigma} n_{\be\sigma'} - \mu \hat{N}.
\end{equation}
As we will see below, this limit captures many important features observed in the more relevant ($J\neq0$) case and allows a simple interpretation of the results. 

\begin{figure}[tb]\center
\includegraphics[width=0.45\textwidth,angle=0,keepaspectratio=true]{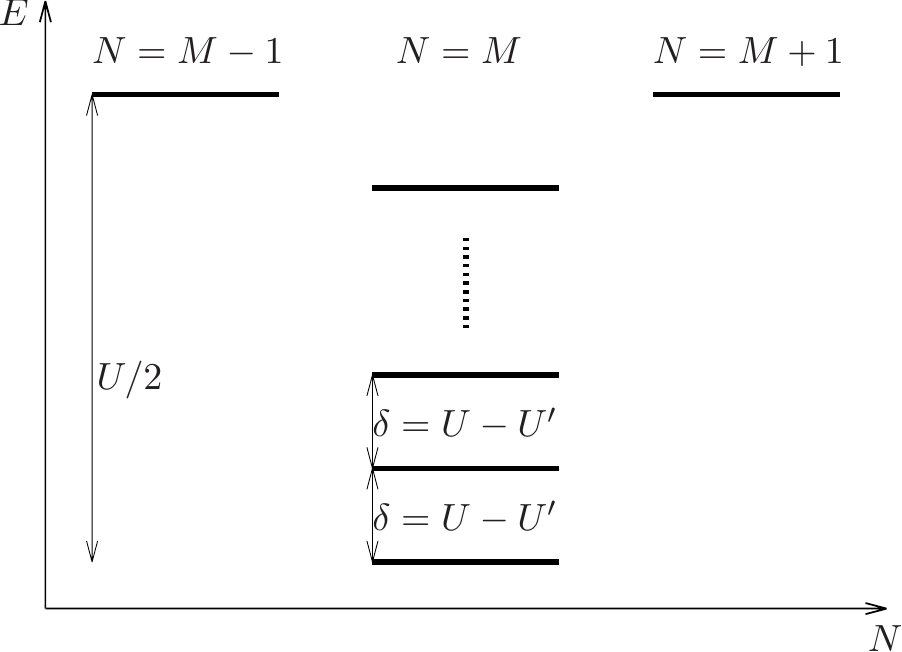}
\caption{Lowest lying multiplet structure of Hamiltonian (\ref{eq:h_up}) at half-filling and $\delta \ll U$. For $U^\prime/U > 1/2$ the lowest lying states in the charge sectors with $N=M-1$ and $N=M+1$ particles are lower in energy than the states in the charge sector with $N=M$ and one or more orbitals having double occupancy.
}
\label{multiplets}
\end{figure}

Figure \ref{multiplets} outlines the lowest lying multiplet structure of $H^{\text{at}}_{J=0}$ for $M$ orbitals and $\mu = U/2+ (M-1) U^\prime$.
The ground state of the atomic Hamiltonian has a level occupation $N=M$ (half-filling) and a degeneracy $(2M)!/M!^2$ for $U^\prime=U$, which is reduced to $2^M$ for $U^\prime<U$. 
The main effect of a finite $U^\prime<U$ is to break the degeneracy of the ground state pushing to higher energies the states on the same charge sector but having one or more orbitals doubly occupied. These states are shifted by $n_d\delta$ where $n_d$ is the number of doubly occupied orbitals and $\delta = U-U^\prime$ (see Fig. \ref{multiplets}).
The gap for charge excitations in the atomic limit is $\Delta(N) = E_0(N+1)+E_0(N-1)-2E_0(N)$, with $E_0(N)$ indicating the energy of the lowest lying state for the charge sector with $N$ electrons. At half-filling $\Delta(M) = U$ is independent of $U^\prime$. 

In the limiting case $U^\prime=0$, $\delta$ is maximal, the orbitals decouple and the model reduces to $M$ copies of the single orbital problem (Hubbard model), which has been extensively studied~\cite{PhysRevLett.69.1236,PhysRevLett.69.1240,PhysRevB.48.7167,PhysRevB.49.10181,PhysRevLett.74.2082,PhysRevLett.82.1915,PhysRevLett.82.4890,bulla1999zero,PhysRevLett.83.3498}. At zero temperature there is a range of values of $U$ where there is a coexistence of metallic and insulating solutions.  The metallic solution has a lower free energy and disappears continuously at a critical interaction $U_{c2}$ leading to a second order phase transition. At finite temperatures, the large degeneracy of the insulating phase leads to a decrease of its free energy due to entropic effects and leads to a first order phase transition at $U=U_c$.

The highly degenerate case for $\delta=0$ ($U^\prime=U$) has also been studied and leads to an enhanced critical interaction \cite{florens2002,florens2004}.
As $\delta$ increases from $0$ to $U$ the charge fluctuations to states having doubly occupied orbitals and $N=M$ are expected to decrease and some questions are in order: What is the role played by excited states with doubly occupied orbitals at the MIT? How does the degeneracy or quasi-degeneracy of the ground state of the atomic Hamiltonian influence the MIT?

\subsection{Two-orbital model}
We first present results for the two-orbital version of the Hamiltonian  of Eq. (\ref{eq:h_up}) at half-filling. The bulk of our analysis is performed using the RISB approach which allows us to explore a wide range of parameters. 
We also perform DMFT calculations using CTQMC at finite temperatures, for specific sets of parameters, which allow us to support our main conclusions.

\subsubsection{RISB results}
\begin{figure}[tb]\center
\includegraphics[width=0.45\textwidth,angle=0,keepaspectratio=true]{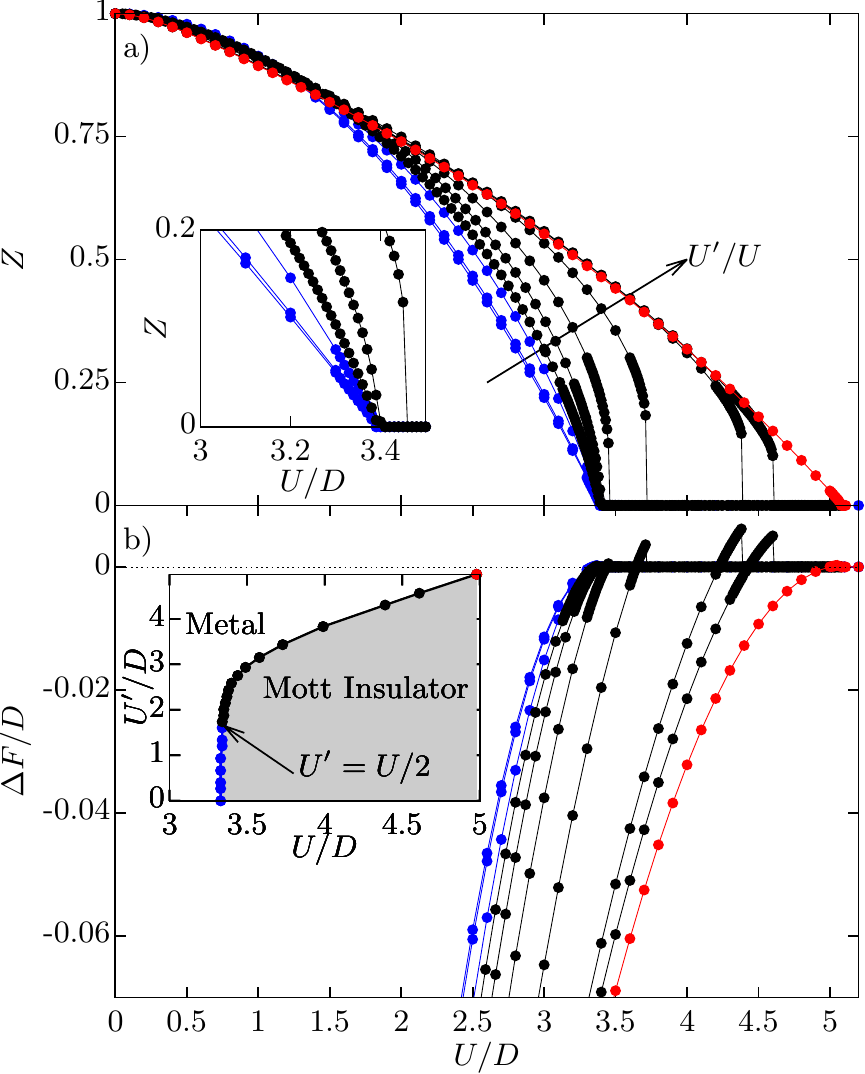}
\caption{(Color online) a) Quasiparticle weight $Z$ vs $U$ obtained using RISB for different values of $U^\prime/U=0,\, 0.2,\, 0.5,\, 0.6,\,0.7, \,0.8,\, 0.9,\,0.98,\,0.99,$ and $1$. The ratio $U^\prime/U$ increases from left to right as indicated in the figure. Blue disks are used for $U^\prime/U\leq 0.5$, black disks for $0.5<U^\prime/U<1$ and red disks for $U^\prime/U=1$.
Inset: Zoom close to $U_{c2}$ for  $U^\prime/U \leq 0.8$.
b) Free energy difference between the metallic and insulating solutions for the parameters of the upper panel. 
Inset: $U$ -- $U^\prime$ phase diagram. 
}
\label{u_up_model}
\end{figure}

Figure \ref{u_up_model} presents zero-temperature results for the quasiparticle weight and the free energy calculated using RISB for $J=0$ and 
fixed values of $U^\prime/U$.
As $U$ is increased,
the quasiparticle weight $Z$ decreases monotonically from its $U=0$ value, $Z=1$.
For values of the interaction $U$ larger than a critical value $U_{c2}$, there is a single solution to the RISB equations which is insulating (Z=0).
The critical interaction remains constant within our numerical precision up to $U^\prime/U\sim 0.5$ and increases monotonously for larger
values of $U^\prime/U$ up to $U^\prime/U=1$ where it attains the maximum value. 

For $U\lesssim 1.5D$, there is a small decrease in $Z$ when $U^\prime/U$ increases as expected from perturbation theory. For larger values of $U$, up to the MIT, an increase in $U^\prime$ enhances $Z$,
i.e., in this regime the interorbital repulsion decreases the effective mass of the electrons \cite{koga2002}.
This behaviour can be traced back to the role of the degeneracy in the auxiliary quantum impurity problem of the DMFT equations: by changing $U^\prime/U$ from 0 to 1, the degeneracy of the associated Kondo model increases from 2 to 6, which leads to an increased Kondo scale \cite{schrieffer1967}. 

The quasiparticle weight Z has two qualitatively different behaviors as $U\to U_{c2}$ depending on the value of $U^\prime/U$. 
For $U^\prime/U=1$ or values of $U^\prime/U$ lower than a critical ratio $\eta_c\sim0.5$, Z vanishes continuously as $U\to U_{c2}$ while for $\eta_c < U^\prime/U <1$ there is a jump in $Z$ from a finite value to zero at $U=U_{c2}$.
These two behaviors are associated with a second and a first order phase transition, respectively. 
To unambiguously characterize the MIT in the different regimes we analyze the behavior of the free energy at the transition which, in the RISB formalism, can be readily evaluated \footnote{In our calculations we used the RISB formalism as an impurity solver in a DMFT scheme. The lattice free energy (Eq. (51) in Ref. \cite{Lechermann2007}) can be written in terms of the impurity free energy (see Ref. \cite{georges1996dynamical})}.
It is important to remark that within the RISB method different choices for the non-interacting density of states $D(\varepsilon)$ lead to different effective bandwidths but do not change the nature of the MIT (see Appendix).

An analysis of the free energy difference $\Delta F$ between the conducting and insulating solutions confirms the conclusions drawn from the analysis of the quasiparticle weight. 
Figure \ref{u_up_model}b) presents $\Delta F$ as a function of the interaction $U$ for different values of the ratio $U^\prime/U$. 
For values of $U^\prime/U$ where $Z$ vanishes continuously at the transition, the slope of $\Delta F$ as a function of $U$ vanishes at $U_c$ (which coincides with $U_{c2}$), while a jump in $Z$ at $U_{c2}$ is associated with a finite value of the slope at $U_c$.
In the inset of Fig. \ref{u_up_model}b) we present the phase diagram in the $U$-$U^\prime$ plane which shows that for $U^\prime/U< \eta_c$ the critical interaction is, within our numerical precision, independent of $U^\prime$ and equal to the single orbital critical $U$.
For larger values of $U^\prime/U$ the critical interaction rapidly increases and attains its maximum value at $U^\prime/U=1$, in agreement with Ref. \cite{koga2002}. 

\begin{figure}[tb]\center
\includegraphics[width=8 cm,angle=0,keepaspectratio=true]{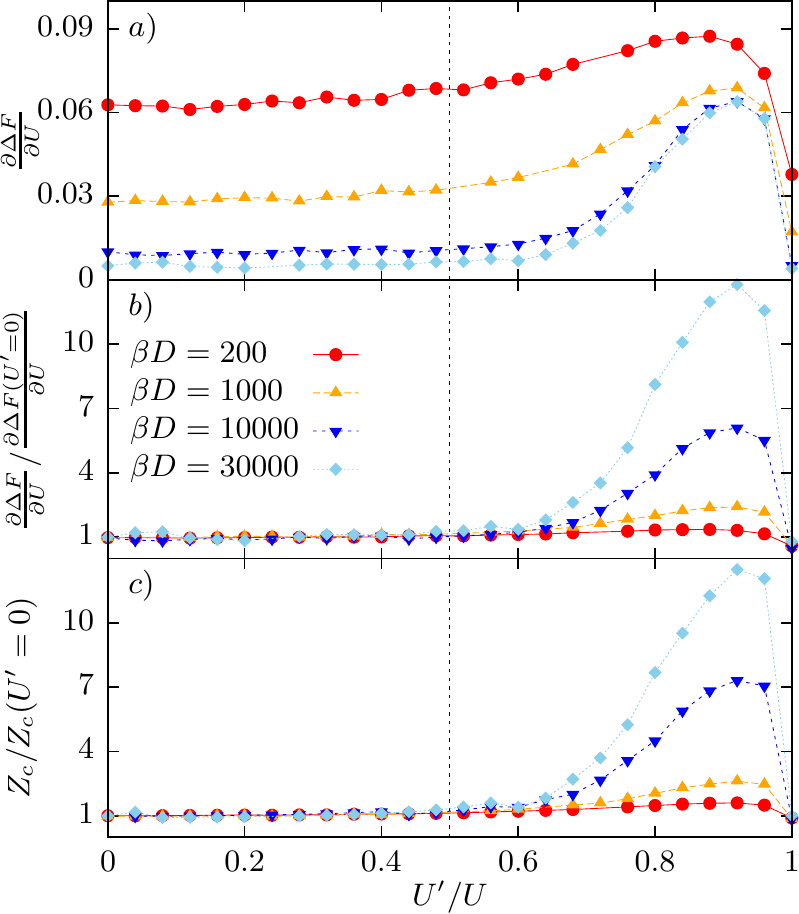}
\caption{(Color online) $a)$ Derivative of the difference between the free energies of the metal and insulator solutions at the free energy crossing point within the RISB method as function of $U^\prime/U$. 
$b)$ Same data scaled with the single orbital temperature dependence.
$c)$ Quasiparticle weight at $U_c$ scaled by its single orbital temperature dependence.
} 
\label{slopes}
\end{figure}

Figure \ref{slopes} $a)$ presents 
the derivative of $\Delta F$ with respect to $U$ at $U_c$ (where $\Delta F=0$) 
as function of $U^\prime/U$ and different temperatures. 
A finite value of this derivative signals a first order transition, which is the case for all the $T > 0$ studied \footnote{In the $U^\prime=0$ case and for high enough temperatures, DMFT calculations indicate that the transition becomes a crossover through a critical end point. This physics is not captured by the slave boson formalism at the mean field level.}. 
For values of $U^\prime/U$ lower than $\sim 0.5$ and for $U^\prime/U=1$, the derivatives decrease with $T$
as in the single orbital case, as can be seen in Figure \ref{slopes} $b)$, where the
data has been scaled with the single orbital temperature dependence.
This indicates that for these values of $U^\prime/U$ the MIT is second order at $T=0$.
For values larger than $\sim0.6$ and smaller than 1, the data no longer follows the single orbital temperature dependence.
The derivatives tends to saturate to a finite value as $T$ is reduced, 
which is consistent with a first order transition at $T=0$.
The strongest first order character is obtained for $U^\prime/U\sim 0.9$ and
both the size of the jump of $Z$ at the transition and the value of the change of the slope of the free energy decrease continuously and 
approach the single orbital dependence as $U^\prime/U$ approaches $1$ or the critical ratio $\eta_c$. 
These considerations place $\eta_c$ between 0.5 and 0.6.
A similar scaling analysis for the value of the quasiparticle weight at the critical value $U_c$, $Z_c$, is shown in Fig. \ref{slopes}$c)$ and leads to the same conclusions.

\begin{table}
	\centering
	\begin{tabularx}{\columnwidth}{@{}c *4{>{\centering\arraybackslash}X}@{}}
		Label&Eigenstates&Occupation&Energy\\
		\hline 
		\hline
		$2PS$ & $|\sigma,\sigma^\prime\rangle$ &2& $-U-U^\prime$\\
		$2PD$ & $|\downarrow\uparrow,0\rangle$, $|0,\uparrow\downarrow\rangle$& $2$ &$-2U^\prime$\\
		$1P$ & $|\sigma,0\rangle$, $|0,\sigma\rangle$ &$1$& $-U/2-U^\prime$\\
		$3P$ & $|\downarrow\uparrow,\sigma\rangle$, $|\sigma,\uparrow\downarrow\rangle$ &$3$&$-U/2-U^\prime$\\
		\hline 
		\hline
	\end{tabularx}
	\caption{Selected eigenstates of the atomic Hamiltonian of Eq. (\ref{eq:h_up}).}
	\label{tab:eigens}
\end{table}

\begin{figure}[tb]\center
\includegraphics[width=8 cm,angle=0,keepaspectratio=true]{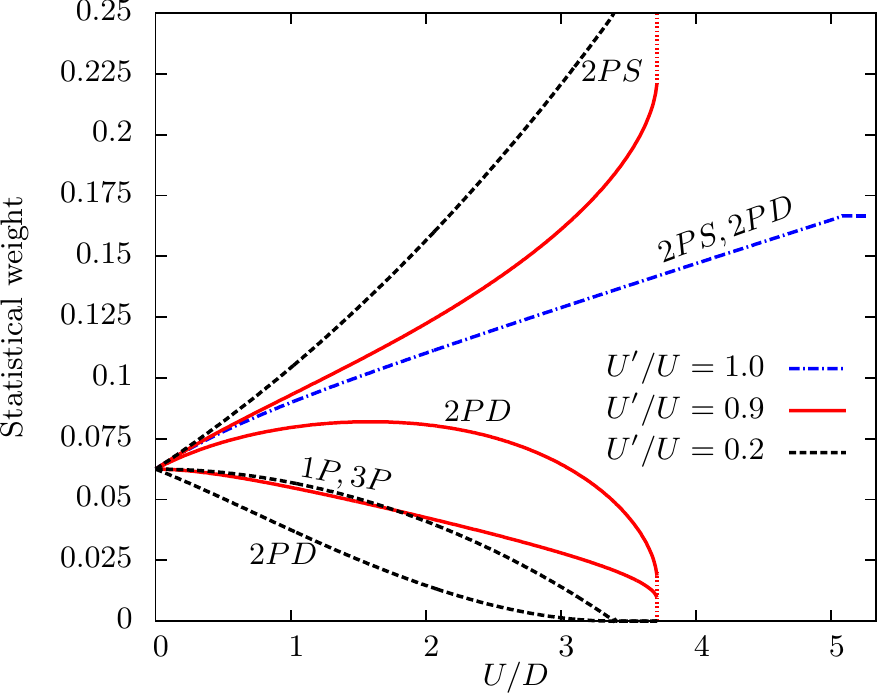}
\caption{(Color online) Total statistical weight of local multiplets in the partition function calculated with the RISB method for three values of the ratio $U^\prime/U$ ($1$, $0.9$, and $0.2$). For $U^\prime/U=1$ the six $2PS$ and $2PD$ states are degenerate and have the same weight. In the insulating phase, within the RISB approximation, there are no charge fluctuations and the full statistical weight is on the ground state sector of the atomic Hamiltonian.}
\label{weightsRISB}
\end{figure}

To gain physical insight into the behavior of the system it proves useful to study 
the statistical weight of the local multiplets (the eigenstates of $H_{J=0}^{\text{at}}$, see table \ref{tab:eigens}) in the partition function.
The most relevant states to be considered in order to understand the physics can be grouped according to their total charge and number of doubly occupied orbitals: states of two electrons
without double occupancy (referred to as $2PS$);
states of two electrons having double occupancy in a single orbital (referred to as $2PD$);
and the single particle states (referred to as $1P$), which due to the electron-hole symmetry considered have the same weight as the
three particle states ($3P$).
Figure \ref{weightsRISB} presents the statistical weight of these states 
calculated within RISB (see Ref. \cite{Ferrero2009,*Ferrero2009a}) for different values of the $U^\prime/U$ ratio.
While the description of the insulating phase by RISB is overly simplified, having a non-zero statistical weight for the ground state only, the overall behavior of the weights is in qualitative agreement with CTQMC results as we show in the next section.
The weight of the $2PS$ states increases as the system approaches the MIT from the metallic side, while the weight of the $1P$ states decreases. 
As expected, for values of $U^\prime/U$ such that the MIT is first order, there is a jump in the weights at the transition that is not present in the other cases at zero-temperature.  
{Besides the jump in the weights, the other feature that makes the $U^\prime/U=0.9$ case different is the behavior of the $2PD$ states.}
The two behaviours emerge from a compromise between the reduction of the kinetic energy and the additional Coulomb repulsion $\delta=U-U^\prime$ associated with  the participation of the $2PD$ states in the ground state wave function.
In the single orbital case ($U^\prime/U = 0$) the weight of the $2PD$ states is continuously and strongly suppressed as $U\to U_{c2}$  \cite{kotliar1984new}. 
For moderated values of $U^\prime/U< 0.5$, the gap $\delta$ between $2PS$ and $2PD$ states is $\sim U$ which leads again to a strong suppression of the weight of $2PD$ states to reduce the Coulomb repulsion as $U\to U_{c2}$. The suppression of the $2PD$ states effectively decouples the two orbitals bringing the system to the single orbital $U^\prime=0$ situation.
This is why increasing $U^\prime/U$ to $0.2$ produces no qualitative change in the behavior of the system close to the Mott transition and $U_{c} \cong U_c^{U^\prime=0}$. 

In the case $U^\prime/U = 1$ and $\delta=0$, the $2PD$ and $2PS$ states are degenerate and have equal statistical weight for all values of $U$. The kinetic energy gain due to charge fluctuations to the $2PD$ states leads to an increase of the critical interaction.
Reducing $U^\prime/U$ from this limit leads to a qualitative change in the behavior of the system.
For all values of $U$ up to the MIT it is convenient to reduce the kinetic energy using the $2PD$ states, which have a significant statistical weight in the free energy. 
As $U$ is increased, however, the low energy quasiparticles are increasingly heavy and the kinetic energy gain relative to the Coulomb energy loss associated with the $2PD$ states $\sim ZD/\delta$ is reduced. 
The MIT transition occurs for the value of $U$ such that it becomes energetically more favorable to suppress the $2PD$ states. This suppression drives the system to the single orbital regime which has a lower critical interaction. Our RISB calculations indicate that this change of regime from a 6-fold degenerate to a 2-fold degenerate Kondo model for the associated quantum impurity problem occurs through a first order phase transition.
This low energy quasiparticle picture obtained with the RISB method is supported at finite temperatures by DMFT calculations using the numerically exact CTQMC as shown in the next section.

\subsubsection{CTQMC results}

We analyzed the MIT at finite temperatures using DMFT with CTQMC as the impurity solver to compare with the RISB results.
\begin{figure}[tb]\center
\includegraphics[width=8 cm,angle=0,keepaspectratio=true]{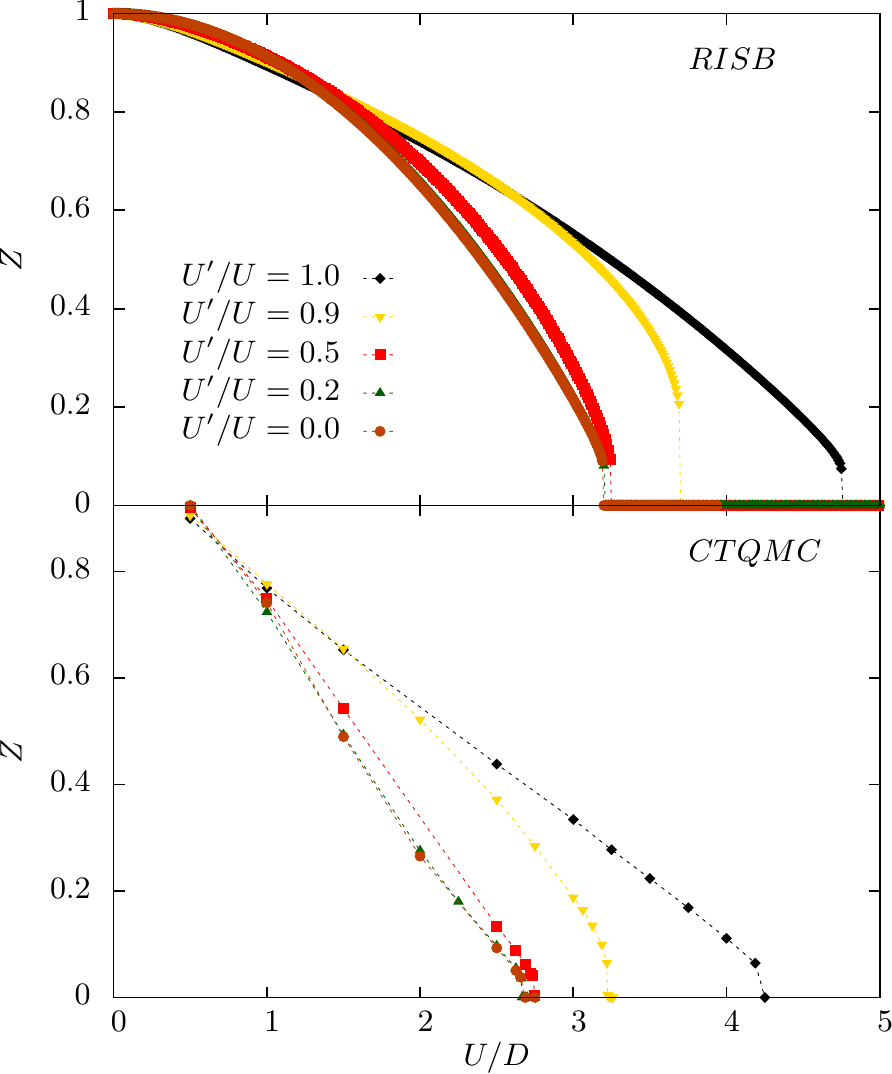}
\caption{(Color online) Quasiparticle weight $Z$ vs $U$ obtained using RISB (top panel) and CTQMC (bottom) techniques as impurity solver for
$\beta D = 200$. 
}
\label{compa_ctqmc}
\end{figure}
Figure \ref{compa_ctqmc} shows the quasiparticle weight as function of $U$ 
for different values of $U^\prime/U$ and $\beta D=200$. Results obtained using the
RISB (CTQMC) technique are shown in the top (bottom) panel.
Both techniques give the same qualitative behavior. 
The main difference is a $\sim20\%$ overestimation of $U_{c2}$ by the RISB method.

\begin{figure}[tb]\center
\includegraphics[width=8 cm,angle=0,keepaspectratio=true]{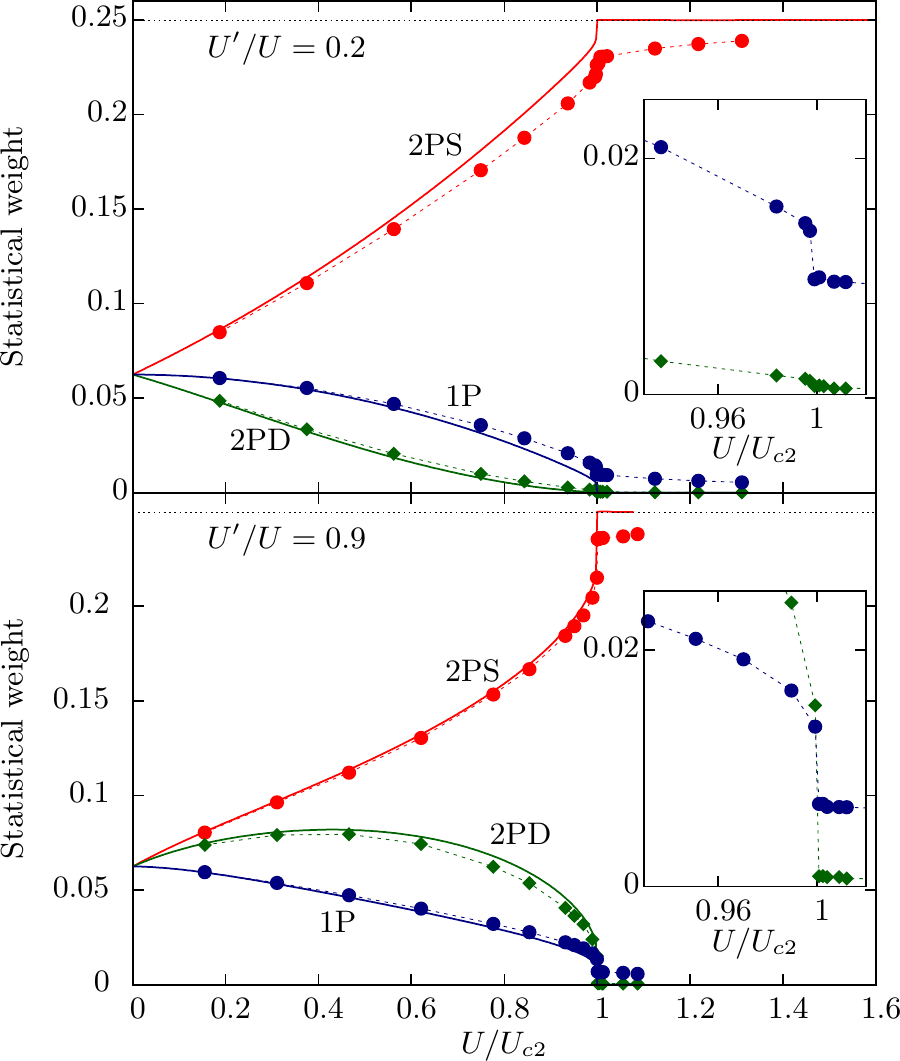}
\caption{(Color online) Statistical weight of local multiplets in the partition function as a function of $U/U_{2c}$ for the two-orbital case and $J=0$. The results were obtained at $\beta D=200$ using RISB (lines) and CTQMC (symbols) as impurity solvers. The insets present in each case
the CTQMC weights close to the transition.} 
\label{weights}
\end{figure}

The statistical weight of the local multiplet states calculated with CTQMC and RISB as quantum impurity solvers is presented in Fig. \ref{weights} as a function of $U$ for two values of the $U^\prime/U$ ratio where a continuous transition ($U^\prime/U=0.2$) and a first order transition ($U^\prime/U=0.9$) are observed at zero temperature. The interaction $U$ is scaled by the critical interaction $U_{c2}$ for each method and value of the ratio $U^\prime/U$ to ease the comparison. 
Both methods present a good quantitative agreement with each other, the largest difference occurs in the insulating phase, 
where the RISB method doesn't describe fluctuations and all the weight is carried by $2PS$ states. 
The agreement between the two methods, in particular concerning the behavior of the statistical weight of the $2PD$ states supports the main conclusions and the interpretation, based on the RISB results, on the previous section. Namely, that there are two markedly different regimes for the behavior of the system at the MIT determined by the ratio $U^\prime/U$. 

The insets in Fig. \ref{weights} present the statistical weight of the excited states $2PD$ and $1P$
as obtained with CTQMC in both phases.
On the metallic side the relative weight of $2PD$ and $1P$ states follow the RISB trend.
On the insulator side, the weight of the $2PD$ states is neglible for any $U^\prime/U$ ratio.
This contributes to obtaining a stronger first-order transition 
for $U^\prime/U = 0.9$ where the $2PD$ states have a much larger weight on the metallic side.

\subsection{Higher orbital number}
\begin{figure}[tb]\center
\includegraphics[width=8 cm,angle=0,keepaspectratio=true]{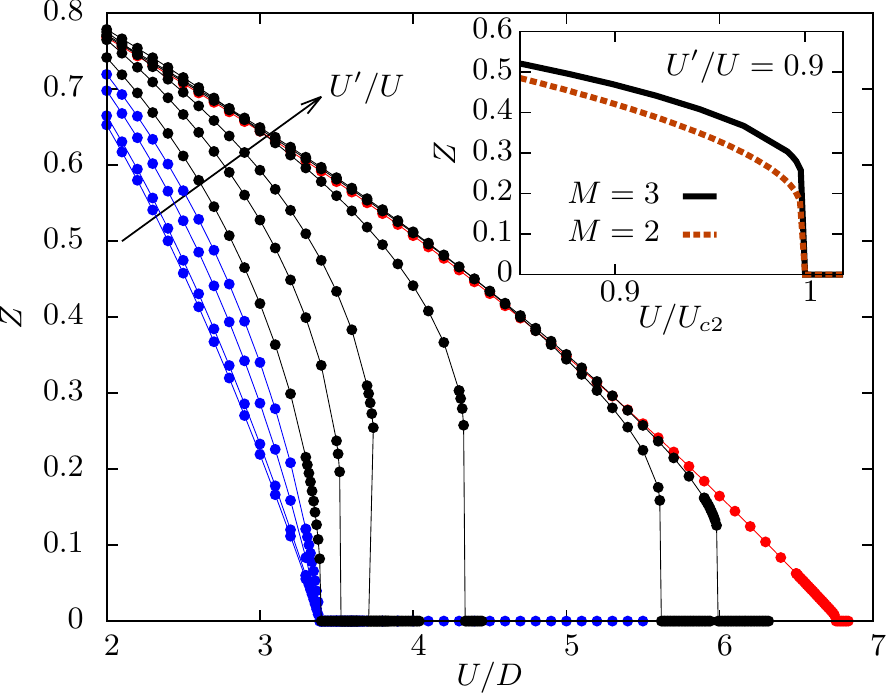}
\caption{ (Color online) Quasiparticle weight for the half-filled three orbital model given by Eq. (\ref{eq:h_up}) and for different values of $U^\prime/U= 0.0,\, 0.2,\,0.4, \,0.5,\,\,0.6,\,\,0.7,\,0.8,\,\ 0.9,\,0.98,\,0.99,$ and $1$.
Inset: quasiparticle weight close to the transition at half-filling for two and three orbitals and $U^\prime/U=0.9$.
}
\label{M3_Upmodel}
\end{figure}

We analyzed the MIT in systems with more than two orbitals in the $J=0$ limit.
Figure \ref{M3_Upmodel} presents RISB results for $Z$ as a function of $U$ for different ratios $U^\prime/U$ in a three-orbital case.  The behavior of $Z$ is qualitatively equivalent to the two orbital case. The main difference is an increase in the critical interaction for all ratios $U^\prime/U> \eta_c$ which is due to the increase in the degeneracy of the ground state of the atomic Hamiltonian that leads to an exponential increase in the Kondo temperature of the associated quantum impurity model. For $U^\prime/U \leq \eta_c$, the critical interaction is, within the numerical precision, the single orbital one, which supports our analysis that in this parameter regime the interorbital correlations are strongly suppressed close to the MIT. Interestingly, the jump in $Z$ increases for a given $U^\prime/U$ ratio with the number of orbitals which can be associated to a stronger reduction in the degeneracy of the ground state for $\delta\neq0$.

For a larger number of orbitals we expect the same pattern to follow: {\it i}) a maximal critical $U$ at the highly symmetric point $U^\prime/U=1$, where the transition is continuous, with a dependence $U_{c2}(M)=U_c(M)\propto M$ with the number of orbitals \cite{florens2002,PhysRevB.87.085115}; {\it ii}) a continuous transition with a single orbital behavior for $U^\prime/U$ lower than a critical ratio $\eta_c\sim 0.5$ which is weakly dependent on the number of orbitals; and {\it iii}) a first order transition, which becomes stronger as the number of orbitals is increased, for $U^\prime/U$ in the interval $(\eta_c,1)$.

\section{Rotationally invariant Kanamori Hamiltonian} \label{sec:model_K}
In the previous section we analyzed the $J=0$ limit in which the atomic gap for charge excitations, 
$\Delta(N)$, depends only on the intraorbital interaction $U$ at half-filling. 
This simplified the analysis of the role played by the interorbital interaction $U^\prime$, since it only changes the structure of the low energy excitations of the atomic Hamiltonian.
In particular we found that two different regimes for the behavior of the statistical weight of the atomic multiplets are obtained depending on the nature of the atomic excitations. Slave boson mean field theory calculations suggest that these two regimes are associated with the order of the MIT at zero temperature and both CTQMC and RISB calculations indicate that they are associated with the strength of the Mott transition at finite temperatures.

In this section we analyze a more physically relevant parameter regime for the Kanamori Hamiltonian using the usual approximation of spherical symmetry for which $U^\prime=U-2J$.
In this case, the interorbital interactions also affect the energy gaps between multiplets in different charge sectors, and the multiplet structure itself is more complex. 

\begin{figure}[tb]\center
\includegraphics[width=8 cm,angle=0,keepaspectratio=true]{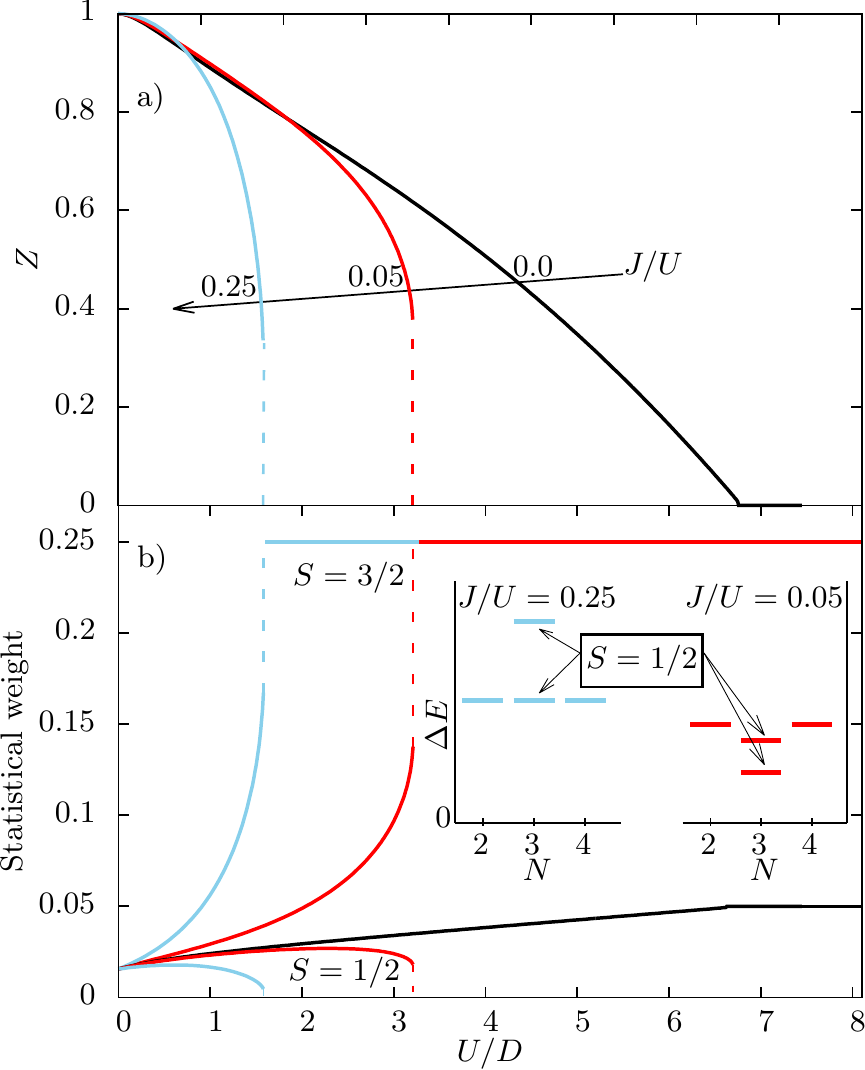}
\caption{(Color online) a) Quasiparticle weight $Z$  vs $U/D$ for different values of $J/U$ in a half-filled three orbital Kanamori system.
	b) Statistical weight for high spin ($S=3/2$) and low spin ($S=1/2$) states. The inset presents the structure of the lowest lying excitations of the atomic Hamiltonian for $J/U=0.25$ and $J/U=0.05$. $\Delta E$ is the energy difference between the lowest lying excited states and the ground state which is in the $N=3$ charge sector and has a spin $S=3/2$.
}
\label{kan_3orb}
\end{figure}
Figure \ref{kan_3orb}a) presents results in a three orbital Kanamori system at half-filling for the behavior of $Z$ as a function of $U$ and different values of $J/U$. As expected \cite{PhysRevLett.107.256401,PhysRevB.83.205112}, the critical interaction is reduced when $J$ is increased due to the increase of the atomic gap for charge excitations $\Delta=U+2J$ at half-filling.
For $J=0$ the model reduces to the $U^\prime=U$ case analyzed in the previous section and the transition is continuous. For the values of $J>0$ studied the transition is first order as signaled by a discontinuity in $Z$ at the MIT. We checked this by analyzing the behavior of the free energy across the transition.

The order of the zero-temperature MIT can, as in the $J=0$ case of the previous Section, be understood analyzing the low lying excitations of the atomic Hamiltonian. In the half-filled case, for the values of $J$ presented in Fig. \ref{kan_3orb}, the charge excitations in the Kanamori model have a higher energy than the low lying excited states in the $N=M$ charge sector. As a consequence, the latter have a finite statistical weight up to the MIT where they are suppressed leading to a first order transition.
Figure \ref{kan_3orb}a) presents the statistical weight of the three particle states with high spin $S=3/2$ ground state and low spin $S=1/2$ excited state. For $J=0$ these states are degenerate and have the same statistical weight for all values of $U$. For a fixed finite value of $J/U$ the energy gap between the atomic ground state and the $S=1/2$ states increases with $U$ and statistical weight is transferred from the latter to the $S=3/2$ states. Confirming this argument we have checked that the critical interaction $U_{c2}$ for finite $J$ is smaller than in the $J=0$ case but larger than in a system with the $S=1/2$ states artificially suppressed. The transition is expected to occur when the gap between the ground state and the $S=1/2$ states is of the order of the effective width of the quasiparticle band $J\sim ZD$.
The strength of the first order transition, as measured by the jump in $Z$ or in the slope of the self-energy, increases with $J$ for $J/U\lesssim 0.1$ and decreases as $J$ increases for $J/U\gtrsim 0.1$. For $J/U\sim 0.3$ the lowest lying charge excitations and spin excitations of the atomic Hamiltonian are nearly degenerate. In this case, we obtain a first order transition, although with a reduced strength compared to the $J/U=0.1$ case. Close to the degeneracy point where the charge excitations and excited states on the charge sector of the ground state have the same energy, we expect the detailed structure of the matrix elements for the coupling of each multiplet with the electron bath and the degeneracy of each multiplet to be important to determine the nature of the transition. This overall behaviour caused by $J$ (first increasing the strength of the MIT and then softening it) is consistent with DMFT results in Ref. \cite{PhysRevB.67.035119,Prushke2005,PhysRevB.83.205112}.

For a filling of a single electron or hole per site, the degeneracy of the ground state multiplet does not depend on the interactions, and the lowest lying excited states are in a different charge sector. Although the value of the critical interaction depends on the number of orbitals and on the value of the interorbital interactions, the transition is always continuous. 

\begin{figure}[tb]\center
\includegraphics[width=8 cm,angle=0,keepaspectratio=true]{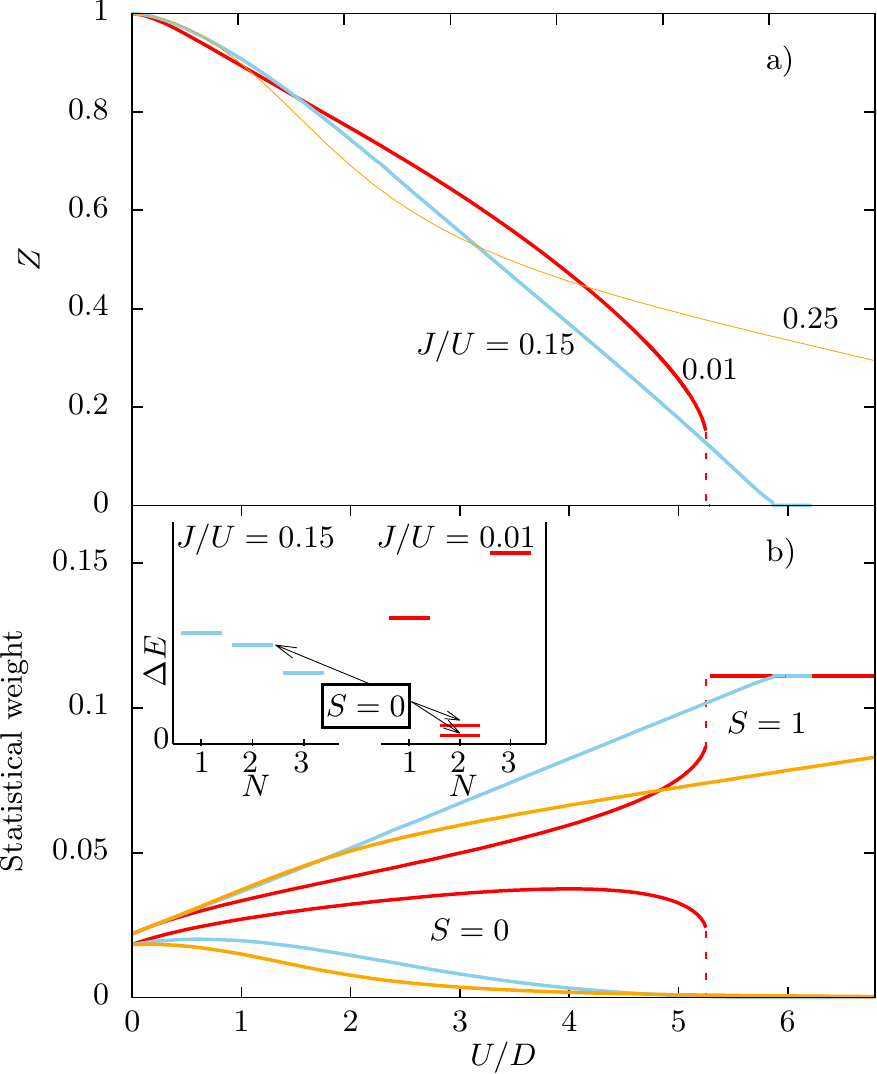}
\caption{(Color online) a) Quasiparticle weight $Z$  vs $U/D$ for different values of $J/U$ in three orbital Kanamori system with an occupation of two electrons per site.
	b) Statistical weight for high spin ($S=1$) and low spin ($S=0$) states. The inset shows the structure of the lowest lying excitations of the atomic Hamiltonian for $J/U=0.15$ and $J/U=0.01$. $\Delta E$ is the energy difference between the lowest lying excited states and the ground state which is in the $N=2$ charge sector and has a spin $S=1$.
}
\label{kan-2}
\end{figure}

For intermediate fillings the situation is more complex but can again be understood in terms of the lowest lying excitations of the atomic Hamiltonian. Figure \ref{kan-2} presents results for the MIT transition in a three orbital Kanamori system with a filling of two electrons per site. 
In this case, the critical interaction has a non-monotonic behavior as a function of $J$ \cite{georges2013}. It decreases for $J/U\lesssim 0.06$ but increases for larger values of $J/U$. The decrease is governed by the breaking of the degeneracy by $J$ of the low energy manifold which also drives the MIT. In the low $J/U$ regime there is a small reduction in the charge excitation gap and for low values of $U$ the behavior of the system closely resembles the $J=0$ case. As $U$ approaches the critical value, the quasiparticle weight decreases linearly as $Z\sim a (U_{c2}^{J=0}-U)/U_{c2}^{J=0}$, where $a\sim 1$ is a constant that depends on the number of orbitals. When the level splitting produced by $J$ becomes of the order of the quasiparticle bandwidth $Z D$ the spin excitations on the $N=2$ charge sector are blocked leading the system to an insulating state at a critical $U<U_{c2}^{J=0}$. Taking a constant ratio $J/U=\alpha$ we have:
\begin{equation}
    U_{c2}(\alpha\sim 0)=\frac{U_{c2}^{J=0}}{1+\frac{\alpha U_{c2}^{J=0}}{a D}},
	\label{eq:Ucalpha}
\end{equation}
which describes accurately the behavior of $U_{c2}$ obtained numerically for $\alpha\ll 1$. These results are consistent with the ones obtained by Attaccalite and Fabrizio \footnote{C. Attaccalite and M. Fabrizio, Phys. Rev. B {\textbf{ 68}}, 155117 (2003). In this work a first order phase transition was obtained using the Gutzwiller approximation in the small J limit (J much smaller than the bare bandwith). Our calculations indicate that the transition continues to be first order even for values of J where the statistical weight of the low lying multiplets is quite different from the $J \to 0$ case.}
using the Gutzwiller approximation, and with CTQMC results at finite temperature \cite{PhysRevB.84.195130}. 

In the regime of large $J$ the charge fluctuations are dominated by the ground state manifold of the $N=2$ charge sector which has $S=1$. 
The MIT is dominated by fluctuations to these states and by the reduced gap for charge excitations in the atomic limit $\Delta^{\text{at}}=U-3J$. 
Figure \ref{kan-2} shows that the quasiparticle weight $Z$ for the ratios $J/U=0.15,0.25$ differs from the small $J/U$ case for moderate values of $U$ where the statistical weight of the low-spin states in the $N=2$ charge sector is strongly suppressed. 
The nature and the critical interaction of the Mott transition is determined by the atomic charge gap that defines a reduced effective interaction $U^{eff}=U-3J$ and by the degeneracy of the ground state manifold. 
In this regime, in which spin excitations on each charge sector are strongly suppressed, the transition is second order at zero temperature and occurs at a critical interaction that can be larger than in the $J=0$ case.

\section{Summary and Conclusions}\label{sec:conclusions}

We analyzed the role played by interorbital interactions on the Mott metal-insulator transition.
To that aim, we performed dynamical mean field theory calculations to treat a model Hamiltonian with Kanamori interactions. 

We first studied, using RISB in the limit $T\to 0$, a simplified case with no Hund's rule coupling ($J=0$) characterized by an intraorbital repulsion $U$ 
and an interorbital repulsion $U^\prime$. 
Depending on the $U^\prime/U$ ratio, we obtained two markedly different regimes characterized by the order of the metal-insulator transition at zero-temperature. 
Remarkably, these regimes are closely associated with the low lying multiplet structure of the atomic Hamiltonian. For $U^\prime/U=1$, the local orbital degeneracy is maximal (all states in the $N=M$ charge sector are degenerate) leading to the highest value of the critical interaction $U_{c2}$ and a second order transition.  
For smaller $U^\prime/U$ ratios, the reduced orbital degeneracy leads to a decrease in $U_{c2}$ and, depending on the value of $U^\prime/U$, to a change in the nature of the transition.
For $U^\prime/U$ in the range $(0.5, 1)$, the lowest-lying atomic excitations have the same charge as the ground state and a significant statistical weight in the metallic phase at the MIT. In the insulating phase the participation of these states is strongly suppressed giving rise to a discontinuity associated with a first order transition for a wide range $(\eta_c,1)$ of values of $U^\prime/U$, with $\eta_c \sim 0.5$.
For lower values of $U^\prime/U$, these states have a higher energy than the charge excitations, and are strongly suppressed in the metallic phase close to the MIT, driving the system to an orbital independent regime and a to continuous phase transition as in the single orbital case.
The results suggest that the nature of the lowest lying excitations of the atomic Hamiltonian determines the order of the transition at zero temperature. We do not expect, however, the critical ratio $\eta_c$ for the change in behavior to be exactly $1/2$ because other factors like the relative degeneracy of the excited states and the intensity of their coupling with the effective bath are likely to play a role in determining its value. In particular we expect the critical ratio to depend (although weakly) on the number of orbitals. 
Moreover, while in the RISB mean field approximation the non-interacting spectral density does not affect the critical ratio $\eta_c$, we do expect it to have an effect on its exact value (see Appendix). 

For the rotationally invariant Kanamori Hamiltonian we find the same approximate connection between the nature of the MIT at zero-temperature and the atomic multiplet structure. 
In this case a finite Hund's rule coupling J reduces the orbital degeneracy favoring high-spin states on each charge sector and also changes the charge excitation gap.
At half-filling, the transition, as predicted by the RISB method, is first order at zero temperature for the range of values of $J$ studied ($0<J/U\leq 0.3$), while in the single electron (or hole) per site case, the transition is second order. 
For intermediate fillings, as two (or four) electrons in three orbitals, the transition is first order for low values of $J/U$, and is second order for $J=0$ and for large enough values of $J/U$.
Similar results have been reported in the literature \cite{Prushke2005,PhysRevB.67.035119}.

As a rule of thumb we find that the low energy multiplet structure of the atomic Hamiltonian, more precisely the nature of the lowest lying excitations, determines the nature of the MIT at zero temperature. 
When the lowest lying excited states are charge excitations we expect a second order transition, but if the lowest lying excited states are on the same charge sector as the ground state, we expect it to be first order. 
Note that this result not only applies to the models studied here but is also consistent with previous reports
where the atomic multiplet structure is changed by the introduction of a crystal field splitting $\Delta$.
For example, in Ref. \cite{PhysRevB.78.045115} it has been found that in a quarter-filled two-orbital system
the order of the transition depends on the magnitude of $\Delta$.
The rule proposed in this work naturally explains this behavior since a large crystal
field splitting implies that the low energy excitations are
charge excitations giving rise to a continuous transition.
For a small crystal field system the lowest lying excitations are
states in which an electron is transferred to the high energy orbital within the same charge sector
and the transition is first order.
A similar behavior is reported in Ref. \cite{PhysRevLett.99.126405} where the effect of a crystal field is analyzed in a two-orbital system at half-filling (see also Ref. \cite{mazza2016field}), and in Ref. \cite{PhysRevB.79.115119} in the three-orbital case.

A detailed analysis using state of the art numerically exact methods~\cite{DMRG2014,NRG3orb2015} would be needed to confirm the RISB results for the nature of the transition at zero temperature.
At finite temperatures, however, the transition is first order and the RISB results are nicely confirmed by the numerically exact continuous-time quantum Monte Carlo. In particular the CTQMC results also show two regimes for the Mott transition characterized by the behavior of the statistical weight of the atomic multiplets, and the strength of the transition as measured by the jump in the quasiparticle weight. 
Remarkably, the first order transition is stronger in the parameter regime where the slave bosons predict a first order transition at zero temperature. 

Materials showing strong first order MITs are known to be good candidates for resistive memory applications \cite{zhou2015mott}. Our results could help as a guide in the quest of this kind of materials. While the MIT at finite but small temperatures is first order, we expect its strength to be determined by the nature of the low lying  excitations of the atomic Hamiltonian.
\acknowledgments
We have profited from discussions with Antoine Georges, Karen Hallberg, Yuriel N\'u\~nez-Fern\'andez, Leonid Pourovskii, Marcelo Rozenberg, and Dieter Vollhardt.

\appendix*
\section{Non-interacting spectral density}\label{ap:lattices}
In the case of no interorbital hybridizations nor orbital asymmetries considered in this paper, it is easy to show that different choices of $D(\varepsilon)$ do not change the nature of the MIT in the RISB approximation. 
Under this approximation, the lattice free energy for the metallic (M) and insulating (I) solutions can be written in the form
\begin{equation}
    F_\text{RISB}^{M(I)}= \bar{\varepsilon} F^{M(I)}(\tilde{u},\tilde{u}^\prime)
    \label{eq:frenesc}
\end{equation}
where, $\tilde{u}=U/\bar{\varepsilon}$, $\tilde{u}^\prime=U^\prime/\bar{\varepsilon}$, 
\begin{equation}
    \bar{\varepsilon}=\int_0^\infty  \epsilon D(\epsilon) d\epsilon,
    \label{eq:avge}
\end{equation}
is the non-interacting average kinetic energy, and  $\{F^{M},F^I\}$ are universal functions.
While the critical interactions and the free energy at the transition do depend on $D(\epsilon)$, the phase diagram is an universal function of $\tilde{u}$ and $\tilde{u}^\prime$. As a consequence, at zero temperature the critical ratio $\tilde{u}/\tilde{u}^\prime= U^\prime/U$ where the transition changes its nature is independent of $D(\varepsilon)$. Although this independence is probably an oversimplification of the RISB approximation, we expect the different regimes to be set primarily by the low energy multiplet structure of the atomic Hamiltonian. 
In particular, to obtain a strong first order transition at finite temperatures we may require, as a rough estimate, the level splitting $\delta=U-U^\prime$ to be smaller than the bandwidth for a range of values of $U>U_c(U^\prime=0)$. This leads to the condition $U^\prime/U> \sim 0.6$ as the critical interation $U_c(U^\prime=0)\sim 3D$ is weakly dependent on the lattice structure \cite{PhysRevB.80.245112}.

\bibliographystyle{apsrev4-1}
\bibliography{references}
\end{document}